\documentclass[twocolumn,aps,prl]{revtex4}

\usepackage{graphicx,color,hyperref}

\usepackage{braket}
\usepackage[utf8]{inputenc}
\usepackage{amsmath}
\usepackage{bbold}

\usepackage{amsfonts}
\usepackage{amssymb}
\usepackage{graphicx}
\usepackage{braket}
\usepackage{natbib}
\usepackage{hyperref}

\begin{document}

\title{
Simulation-Free Fidelity Estimation via Quantum Output Order Statistics
}

\author{ 
Tobias Micklitz 
}
\affiliation{Centro Brasileiro de Pesquisas F\'isicas, Rua Xavier Sigaud 150, 22290-180, Rio de Janeiro, Brazil }

\date{\today}

\begin{abstract}
We introduce a simulation-free method to estimate the fidelity of large quantum circuits 
based on the order statistics of measured output probabilities from highly entangled, chaotic states. 
The approach requires only the highest-probability output bitstrings---the most frequently observed 
measurement outcomes---and builds on exact analytical results for the order statistics of Haar-random quantum states
derived here. Analyzing their modification under depolarizing noise,  
we propose a scalable fidelity estimator, validated on Google's 12-qubit Sycamore experiment 
and further supported by numerical simulations. We demonstrate its practicality for 
intermediate-scale quantum circuits, where cross-entropy 
benchmarking is costly and direct fidelity estimation is difficult. 
\end{abstract}

\maketitle

{\it Introduction:---}Universal quantum processors are approaching 
the regime where classical simulation becomes infeasible, marking the onset of 
quantum supremacy~\cite{Preskill2012,Arute2019}. This raises a fundamental challenge: 
how can we verify the correctness 
of quantum computations or benchmark their performance 
when the ideal output distribution cannot be efficiently simulated?

Fidelity estimation methods, such as quantum state tomography 
and direct fidelity estimation, work well for small or structured systems 
but do not scale beyond $\sim20$ qubits~\cite{Rodriguez2025, Wang_2025, Zhang2021, daSilva2011}. 
Cross-entropy benchmarking, used in early quantum supremacy experiments, 
relies on classical simulation of 
ideal outputs and becomes computationally demanding for circuits larger than $\sim30$ 
qubits. 
This motivates the search for simulation-free fidelity estimation techniques that 
remain practical in the regime of $\gtrsim {\cal O}(20)$ 
qubits, where classical 
simulation becomes increasingly expensive, and direct fidelity estimation is often 
infeasible.

\begin{figure}[t!]
\centering
\vspace{.3cm}
\includegraphics[width=8.4cm]{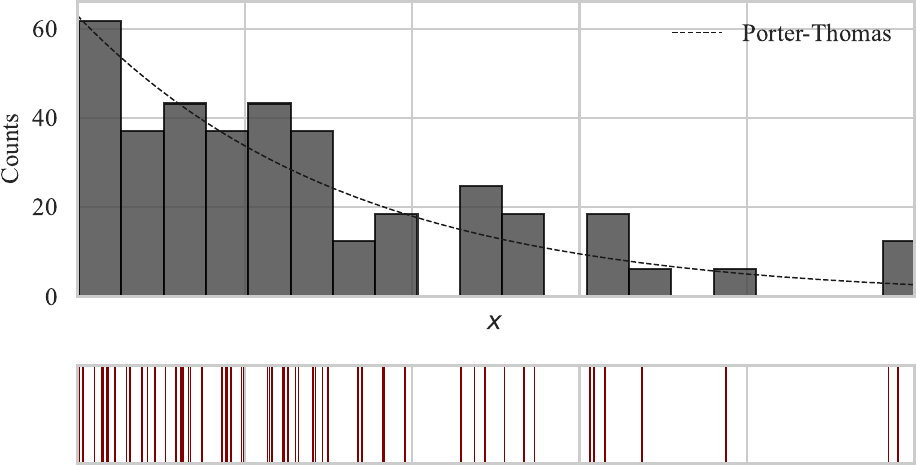}
\caption{
Histogram of bitstring probabilities from a single Haar-random quantum state of 
$N=6$ qubits ($D=64$), compared to the finite-$D$ Porter-Thomas distribution 
(dashed line). Fluctuations resemble those of optical speckle, with each realization 
characterized by a unique fingerprint of ordered bitstring probabilities $p_k$, 
shown in the bottom panel. 
}
\label{fig1}
\end{figure}

Universal quantum computers can, by definition, implement arbitrary unitary operations. 
Among these are chaotic circuits drawn from the Haar measure that produce 
 highly entangled states with universal statistical features. 
Notably, the output probabilities follow the Porter-Thomas distribution, which has been 
leveraged to detect noise and benchmark experimental implementations~\cite{Magesan2010, Boixo2017, Choi2023, Elben2023}. 
However, experimental access provides only partial information about the output distribution. 
Each projective measurement yields a single bitstring, and its probability  
must be estimated from repeated measurements. The resulting distribution 
forms a quantum speckle pattern---a distinctive fingerprint reflecting 
the interference structure of the quantum state~\cite{Boixo2017, Aaronson2017, Bouland2018, Mark2023} 
(see also Fig.~\ref{fig1}). While most of the probability is spread over exponentially 
many low-probability outcomes, the most frequently observed bitstrings carry significant 
information about the state~\cite{Neill2018, Villalonga2022}. 
In large scale experiments, such as recent supremacy demonstrations~\cite{Arute2019, Zhong2020}, 
only the top-ranked outcomes are experimentally accessible, yet this partial 
fingerprint encodes valuable signatures of fidelity.

Inspired by the statistical structure of chaotic quantum circuits, 
we here develop a simulation-free fidelity estimator based on the order 
statistics of measured output probabilities. We derive exact analytical expressions for the 
expected order statistics of Haar-random states and characterize their distortion under depolarizing 
noise. This leads to a maximum likelihood estimator that infers circuit fidelity from only a small number  
of top-ranked measurement outcomes. Validated on Google's 12-qubit Sycamore data and supported by 
numerical simulations, the method offers a practical tool for fidelity estimation 
in intermediate-scale universal quantum processors.

\begin{figure*}[t!]
\centering
\includegraphics[width=17cm]{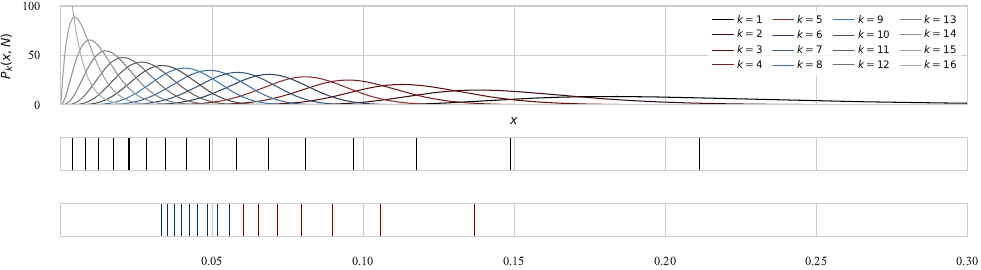}
\vspace{-.24cm}
\caption{
    Upper panel: Decomposition of Porter Thomas distribution for $N=4$ qubits ($D=16$) 
into its ordered statistical components. Distributions $P_k(x,N)$ transition from Gumbel-like 
to Gaussian and approximately exponential profiles as the order increases from the largest 
($k=1$) to the smallest ($k=16$) probability. Middle panel: Ideal 'fingerprint' of ordered 
bitstring probabilities $p_k$, constructed from the first moments of the order statistics 
Eq.~\eqref{eq:order_statistics}, $\langle p_k \rangle = \frac{1}{D} [\psi(D+1) - \psi(k)]$, 
where $\psi(k)$ is the digamma function. Notice the characteristic spacing between successive values, 
$\langle p_k\rangle - \langle p_{k-1} \rangle = 1/k$, reflecting the logarithmic structure of the 
ranked probabilities. Bottom panel: The same fingerprint in the presence of homogeneous depolarizing 
noise Eq.~\eqref{eq:noise_model} with fidelity $f=0.5$, similar to that reported in the $12$-qubit 
Sycamore experiment~\cite{Arute2019}. Probabilities larger than the uniform value $1/D$ (red) are 
compressed by $f<1$, while those smaller (blue) are inflated toward it.}
\label{fig2}
\end{figure*}

{\it Order Statistics of Haar-Random States:---}Consider a fully chaotic quantum 
 state of $N$ qubits $|\psi\rangle = \sum_{k=1}^D z_k|k\rangle$, 
 with $D = 2^N$, and output probabilities $p_k=|z_k|^2$ defined 
in the computational basis. 
Assuming no symmetries, 
the joint distribution of these probabilities 
is uniform over the $D$-dimensional simplex,
 constrained 
only by normalization,
$P_{\rm CUE}(p_1,p_2,...,p_D) 
=
(D-1)!
\delta\left( 
p_1+p_2+...+p_D -1
\right)$.   
The marginal distribution of individual 
$p_k$-values is the Porter-Thomas distribution, 
$P_{\rm PT}(x) 
= 
(D-1)(1 - x)^{D - 2}$, 
characteristic of the Circular Unitary Ensemble (CUE)~\cite{Haake2010}. 
While the support of the PT-distribution becomes exponentially 
large for increasing $N$, making it experimentally 
 inaccessible, the top-ranked measurement outcomes remain 
 observable and statistically robust.  
This motivates a decomposition, 
$DP_{\rm PT}(x)=\sum_{k=1}^D P_k(x,N)$, 
into order statistics, 
where we define the normalized 
probability distribution for the $k$-th 
largest outcome to take value $x$,
\begin{align}
\label{eq:order_statistics_0}
P_k(x,N)
&\propto 
\int_{S_x} dX\,
 P_{\rm CUE}(p_1,..,p_k=x,...,p_D),
\end{align}
with integration domain
$S_x =[x,1]^{k-1}\oplus [0,x]^{D-k}$  
enforcing the ordering condition. 

Expressing the $\delta$-function as an auxiliary integral, 
the integral over $S_x$ can be evaluated exactly,  
yielding the closed-form expression~\cite{suppmat},  
\begin{align}
    \label{eq:order_statistics}
P_k(x,N) 
&=
{\cal N}
\sum_{j=k}^{j_{\rm max}}
\begin{pmatrix}
D-k\\ j-k
\end{pmatrix}
(-1)^j (1 - jx)^{D-2},
\hspace{-.05cm}
\end{align} 
where $j_{\rm max} 
= \min\left( D, 
\left\lfloor 1/x 
\right\rfloor 
\right)$, with $[\ldots]$ the floor function, and ${\cal N}$ the 
normalization constant~\cite{footnote4}. 
Eq.~\eqref{eq:order_statistics} is the central technical result of this work. 
The case $k=1$ was previously derived in Ref.~\cite{Lakshminarayan2008},
and extending it to all $k$ reveals a crossover across the 
rank spectrum, see also Fig.~\ref{fig2}.

For small ranks
$k\ll D$, the order statistics describe the extreme-value tail of 
the PT distribution, yielding a Gumbel-like form~\cite{Coles2001}, 
with mean scaling as
$\ln(D/k)/D$ and variance $\sim1/D^2$. 
Near the median rank $k \sim D/2$, the distribution 
transitions to an approximately Gaussian form, 
consistent with the central limit theorem. 
In this regime, 
the mean scales as $\sim 1/D$ and variance as $\sim 1/D^3$. Finally,  
for ranks near $k\lesssim D$, the distribution becomes sharply peaked 
near zero with suppressed mean $\sim 1/D^2$ and variance $\sim 1/D^4$.  
In particular, the distinctive statistical structure of the top-ranked 
outcomes offers a simulation-free framework for fidelity estimation 
based on Eq.~\eqref{eq:order_statistics}.

{\it Noise Model and Likelihood Function:---}The dominant source of error in chaotic quantum circuits 
is often captured by a uniform depolarizing noise channel. It 
 closely approximates average error behavior 
in large-scale circuits, especially when randomized compiling or Pauli twirling are  
applied~\cite{Hashim2020,Arute2019,Wallman2016,Kandala2019,Boixo2018,Emerson2005}, 
and models decoherence using a single parameter, the fidelity 
$f$~\cite{footnote3}. 
Under depolarizing noise, each ideal output probability 
$p_k$ is affinely shifted toward the uniform value 
$1/D$, the median of the Porter-Thomas distribution
\begin{align}
\label{eq:noise_model}
p_k(f) = f p_k + (1 - f)/D,
\end{align}
where $p_k(f)$ is the noisy output probability. 

Practically, Eq.~\eqref{eq:noise_model} 
suppresses statistical characteristics of ideal chaotic quantum circuits. 
Probabilities larger than $p_k$ are suppressed toward 
the fixed point 
$1/D$, and smaller ones are increased, 
without altering their ordering, see also Fig.~\ref{fig2}. 
The deformation is most pronounced for the 
most probable bitstrings that dominate sampling, as they appear most frequently. 
The effect of depolarizing noise on order statistics is captured 
by a simple rescaling of the ideal Haar distribution
\begin{align}
    \label{eq:order_statistics_noise}
P_k(x; N, f) 
&= 
P_k\left( x_f, N \right), \quad 
fx_f 
=  
x - (1 - f)/D. 
\end{align}
Since each rank shifts differently, yet in a manner fully determined by the single fidelity parameter 
$f$, Eq.~\eqref{eq:order_statistics_noise} enables rank-resolved fidelity estimation and 
provides a nontrivial test for the noise structure.

Building on this rescaling, we extract fidelity from experimental data via order 
statistics by constructing a likelihood over the observed ranks. For each circuit realization 
$m = 1, \dots, M$, let 
$\{p_k^m\}_{k=1}^K$ denote the $K$ largest measured probabilities, normalized per realization. 
Assuming uniform depolarizing noise, we introduce the likelihood function 
\begin{align}
\label{eq:log_likelihood}
\ln \Lambda(f; N, \{p_k^m\}) 
= 
\sum_{m=1}^M \sum_{k \in K^*} \ln P_k(p_k^m; N, f),
\end{align}
where $P_k(x; N, f)$ is the noise-deformed Haar distribution 
from Eq.~\eqref{eq:order_statistics_noise}, and $K^* \subseteq \{1, \dots, K\}$ 
is a selected subset of the measured ranks. 
The fidelity estimate is obtained by maximizing $\Lambda(f)$ 
over $f \in [0, 1]$.  

Eq.~\eqref{eq:log_likelihood} supports both, sampling across different circuit realizations 
at a single fixed rank and single-circuit inference averaging over multiple ranks, 
which is especially useful when only limited data is available. 
Correlations between ranks within a single realization 
introduce bias in the likelihood, 
but this bias weakens with increasing circuit size. 
Further mitigation is possible by sampling well-separated ranks 
or combining disjoint subsets.
In the following, we focus on ranks up to $K\sim {\cal O}(N)$, 
which lie deep within the Gumbel-like tail of the distribution and carry 
high sensitivity to fidelity, offering informative and simulation-free 
diagnostics for large quantum circuits.

\begin{figure}[t!]
\centering
\vspace{.1cm}
\hspace{-.6cm}
\includegraphics[width=9cm]{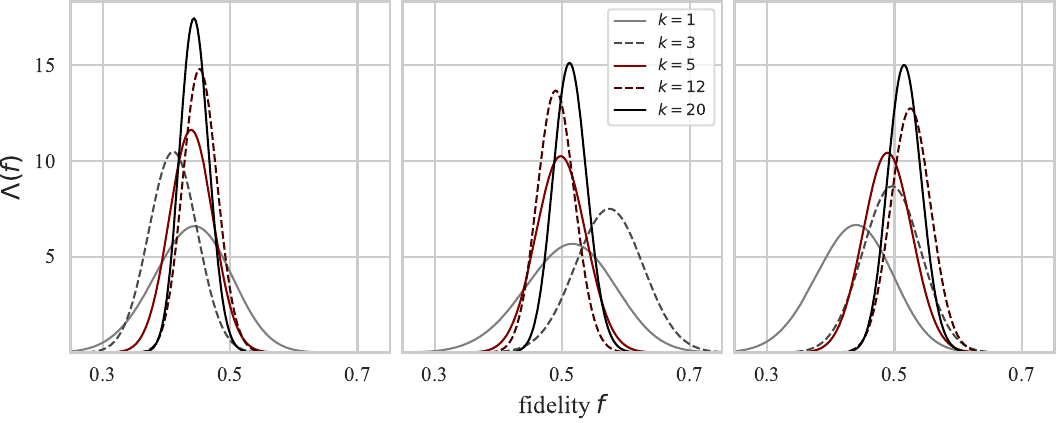}
\caption{Illustrative example of the likelihood function $\Lambda(f)$ vs. fidelity $f$ 
for a single circuit realization ($N=12$) from Google's Sycamore experiment, evaluated 
at different ranks $k-1,3,5,12,20$. The approximately Gaussian shape near the peak, 
the narrowing of the distribution with increasing $k$, and the rank-consistent location 
of the maxima exemplify generic features of the likelihood 
function across system sizes. These features underpin the rank-averaging strategy used 
in our estimator and hold more generally beyond the $N=12$ case shown here.
}
\label{fig3}
\end{figure}

{\it Validation and Scalability:---}We validate our fidelity estimation using public data from Google's 12-qubit 
Sycamore experiment~\cite{Arute2019}, analyzing 20 circuit realizations with $5\times 10^5$ measurements
each. 
Although this system size allows near-complete sampling, we restrict our analysis to only the 
${\cal O}(10)$ most frequent bitstrings per circuit, in line with our focus on scalable, low-rank diagnostics. 
Fig.~\ref{fig3} illustrates typical features of the likelihood function $\Lambda(f)$ 
for different ranks $k$ within a single circuit. Even at this moderate $N$, the likelihoods 
show approximately Gaussian profiles, consistent peak locations across ranks, and narrowing widths 
with increasing $k$. These generic features persist at larger $N$ and underpin the effectiveness 
of rank-averaging strategies.

We test the method using two averaging strategies~\cite{suppmat}: (i) across 
circuits at fixed rank, and (ii) across ranks within individual circuits. 
Both strategies yield fidelity estimates in the range 
$f\simeq 0.45-0.50$, consistent with cross-entropy benchmarking~\cite{Arute2019}, 
and indicate robustness to subsampling and circuit-to-circuit 
fluctuations under a global noise model. 
This highlights a core feature of the method: it requires only a few circuit 
realizations and a small number of top-ranked outcomes, avoiding classical simulation 
or complete output sampling.
For larger systems, 
direct evaluation of Eq.~\eqref{eq:order_statistics} 
becomes numerically unstable due to oscillatory cancellations among large binomial 
terms. To enable stable and scalable estimation, we use a low-rank, large-$D$ 
approximation~\cite{footnote2},  
\begin{align}
P_k(x, D) 
&= 
\mathcal{N} \, e^{-k(D-2)x} \left(1 - e^{-(D-2)x}\right)^{D-k},
\end{align}
where $\mathcal{N}$ ensures proper normalization~\cite{footnote5}.
This approximation preserves key statistical properties 
(mean, variance, peak location) up to $\mathcal{O}(1/D)$ 
corrections and is used in the Sycamore analysis.
It allows for a straightforward estimate of 
the width of the likelihood function, scaling as $\sim f^2/(\sqrt{k} N)$. 
Fidelity estimates thus become increasingly sharp with system size, as we also verified 
in simulations (see Ref.~\cite{suppmat}). 
 This favorable scaling supports applicability to quantum advantage experiments, 
though precision remains ultimately limited by finite sampling, especially as $N$ increases.

{\it Finite Sampling and Estimator Robustness:---}A major practical challenge in fidelity 
estimation is the limited number of measurement shots available per circuit realization. 
Empirical output probabilities are obtained as $p_k=n_k/S$, where $n_k$ is the observed 
count of the $k^{\rm th}$ most frequent bitstring and $S$ is the total number of shots. 
In large circuits, relevant probabilities scale as $p_k \sim \ln D/D$. Achieving a 
relative accuracy $\varepsilon_{\mathrm{rel}}$ in these estimates then requires a 
shot count scaling as 
\begin{align}
    \label{eq:estimate_shots}
    S \gtrsim \frac{2^N}{\varepsilon_{\mathrm{rel}}^2 N},
\end{align}
which quickly becomes experimentally prohibitive. 
Beyond this threshold, shot noise dominates the signal, making the probability-based 
likelihood estimation unreliable.

To overcome this limitation, we consider the 
extension of Eq.~\eqref{eq:log_likelihood} to a count-based likelihood estimator,  
which works directly with raw measurement counts. 
This avoids the need to explicitly construct empirical probabilities by using that 
observed counts $n_k$ across $S$ shots follow a binomial distribution,  
$P(n_k,S)=\binom{S}{n_k}p^{n_k}_k(f)\left(1-p_k(f)\right)^{S-n_k}$. 
The fidelity-dependent probabilities  
$p_k(f)$ are now derived from the noise-deformed Haar statistics in  
Eq.~\eqref{eq:order_statistics_noise}~\cite{footnote6}.

In the limit of large shot-numbers $S\gg1$, the binomial distribution 
simplifies to a Poisson distribution, motivating 
the (unnormalized) log-likelihood function for  
observed counts
\begin{align}
\label{eq:log_likelihood_count}
    \ln \Lambda(f; N, \{n_k\}) 
    &= 
    \sum_{k \in K^*} \left( n_k \ln p_k(f) - S p_k(f) \right).
\end{align}
Here, $f$-independent terms are omitted, 
and fidelity is estimated by maximizing $\ln \Lambda$,
as before.
This count-based  estimator 
is based on minimal modeling assumptions, and remains remarkably 
robust even under severe shot limitations. 
Its width scales as $\sim \sqrt{fD/(KS\ln D)}$, 
improving over the bound in Eq.~\eqref{eq:estimate_shots} 
 by a factor $f/K$ ($K$ is the number of ordered ranks kept 
 in Eq.~\eqref{eq:log_likelihood_count}), which   
  allows for reliable fidelity estimation 
 well below the naive shot threshold.

\begin{figure}[t!]
\centering
\includegraphics[width=8.2cm]{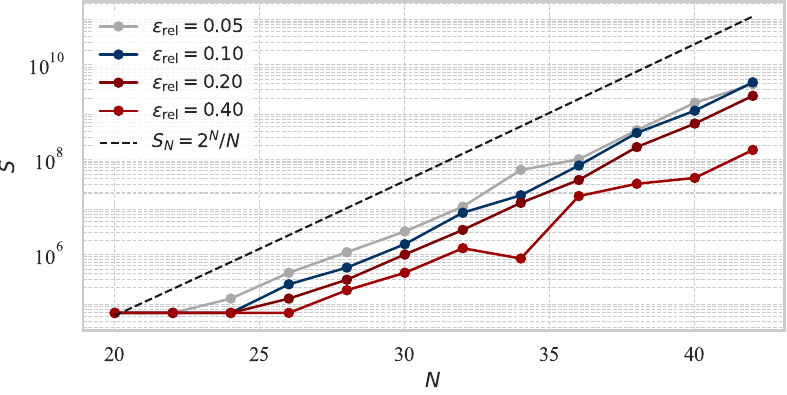}
\vspace{-.24cm}
\caption{
Number of shots $S$ required for system size $N$ 
to correctly estimate the fidelity $f = 0.10$ within relative error 
thresholds $\varepsilon_{\mathrm{rel}}=0.05,0.1,0.2,0.4$. 
The dashed black line shows the reference scaling $S_N = 2^N/N$. 
Results are averaged over $M = 1000$ realizations ($M=2000$ for 
$\varepsilon_{\mathrm{rel}}=0.4$). 
}
\label{fig4}
\end{figure}

\begin{figure}[t!]
\centering
\includegraphics[width=8.2cm]{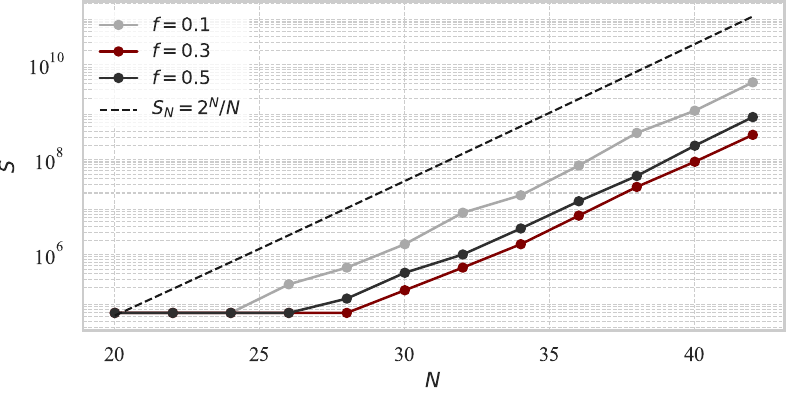}
\vspace{-.24cm}
\caption{
Number of shots $S$ required as a function of system size $N$ to estimate 
fidelity within a fixed relative error threshold $\varepsilon_{\mathrm{rel}} = 0.10$, 
for different fidelities $f= 0.1, 0.3, 0.5$. 
The dashed black line indicates the naive scaling $S_N = 2^N/N$. 
Results are averaged over $M = 1000$ realizations. 
}
\label{fig5}
\end{figure}

 To validate the scalability of Eq.~\eqref{eq:log_likelihood_count}, 
 we simulate chaotic circuits of $N = 20-42$ qubits, 
 each subject to homogeneous depolarizing noise of known strength. 
 Figs.~\ref{fig4} and~\ref{fig5} show the 
 number of shots required to correctly estimate the fidelity within a 
 given relative error threshold for different thresholds and fidelities, 
 respectively. 
Notably, accurate fidelity estimates are 
obtained even when  
$S$ is well below the minimum threshold in Eq.~\eqref{eq:estimate_shots}, 
$S\gtrsim 2^N/N$. 
This robustness underscores the practical viability of the method, particularly 
where the probability-based estimator Eq.~\eqref{eq:log_likelihood} fails. 
The effect is reminiscent of the birthday paradox: despite the exponentially large output space, 
the most probable bitstrings recur often enough to carry meaningful 
information---even under limited sampling.

In the simulations, we use $K=500$ ranked outcomes to keep computational cost 
moderate. In single instances we verified that doubling the 
number of included ranks $K$ consistently reduces the required shot count $S$ by 
approximately a factor of two. This suggests room for improvement through rank 
optimization, as well as more advanced techniques such as higher-order statistics 
or Bayesian inference, all of which lie beyond the scope of this work.

{\it Summary and Conclusion:---}We have introduced a simulation-free method 
to estimate fidelity in chaotic, universal quantum circuits, 
based on the order statistics of top-ranked measurement outcomes. 
By focusing on low-rank statistics and employing a simple depolarizing noise model, 
our approach circumvents the exponential cost of classical amplitude calculations, 
enabling fidelity estimation beyond the reach of straightforward classical simulations. 
Validation on Google’s 12-qubit Sycamore data shows close agreement 
with cross-entropy benchmarking, 
demonstrating robustness across circuits and rank selections. 
The rich statistical structure of the order statistics produces 
 well-resolved likelihood peaks, enabling 
accurate fidelity estimates from modest measurement data.
To overcome limitations from finite sampling, 
we generalized the method to a frequency-based estimator that works 
directly with raw measurement counts. 
This estimator remains accurate even under severe shot constraints, 
extending fidelity estimation to larger systems  
where the probability-based method becomes unreliable.

Though simple and grounded in minimal assumptions, the method’s efectiveness underscores 
the power of low-rank order statistics for simulation-free quantum diagnostics. 
The core idea introduced here opens avenues for further refinement and optimization.  
More broadly, order statistics offer a versatile framework to probe 
the structure and complexity of quantum output distributions. 
Recent works~\cite{Vallejos2021,Tacla2023} explore related diagnostics 
based on Lorenz curves and majorization, suggesting deeper insights into 
quantum amplitude fluctuations. Our approach complements these efforts 
by providing a practical tool for validating 
intermediate-scale universal quantum processors.

{\it Acknowledgments:---}We thank F.~de~Melo, R.~O.~Vallejos, A.~B.~Tacla, F.~Monteiro, J.~Telles de Miranda, 
A.~Altland, and M.~Micklitz for fruitful discussions, and Brazilian agencies CNPq and FAPERJ for financial support. 



\begin{thebibliography}{33}
\bibitem[{foo({\natexlab{e}})}]{footnote6}
\bibinfo{note}{That is, $p_k= (\psi(D+1) - \psi(k))/D$, with $\psi(k)$ the
  digamma function.}
\end{thebibliography}

\newpage

\begin{appendix}

\section{Order statistic}

To calculate the order statistics of outcome probabilities  
$p_k\equiv|\langle k |\psi\rangle|^2$ of a Haar 
random quantum state of $N$ qubits,
we start out from the probability distribution 
$P_k(x)\equiv P_k(x,N)$, 
for the $k^{\rm th}$ largest probability to take
the value $x$,   
\begin{align}
\label{F}
P_k(x)
&\propto
\int_{S_x} dX\,
P_{\rm CUE}(p_1,..,p_k=x,...,p_D),
\end{align}
where 
$P_{\rm CUE}(p_1,...,p_D) 
\propto
\delta\left( 
p_1+p_2+...+p_D -1
\right)$, 
and the 
integral is over the restricted simplex 
$S_x =[x,1]^{k-1}\oplus [0,x]^{D-k}$. 
That is,
\begin{align}
\label{app_F}
P_k(x)
&\propto
\prod_{i=1}^{k-1} 
\int_x^1 dp_i 
\prod_{j=k}^{D-1} 
\int_0^x dp_j\,
\delta_{X + x - 1},
\end{align}
where  $X= \sum_{i=1}^{D-1}p_i$. 
Expressing the $\delta$-function in terms of an auxiliary integral
\begin{align}
\delta_{X + x - 1}
&=
\int_{-\infty}^\infty 
\frac{d\xi}{2\pi}
e^{i\xi^+(p_1+...+p_{D-1}+x-1)},
\end{align}
with $\xi^+=\xi + i0$ to guarantee convergence, 
we 
extend the upper  boundary in $p_i$-integrations 
to $\infty$
and perform integrations over $p_i$, $p_j$ 
to arrive at
\begin{align}
P_k(x)
&\propto
\int
\frac{d\xi}{ 2\pi} \, e^{i\xi^+(x-1)}
\left( { e^{i\xi^+ x} - 1 \over i\xi^+}\right)^{D-k}
\left( {ie^{i\xi^+ x} \over \xi^+} \right)^{k-1}.
\end{align}
Expanding the first monomial 
in the integrand in a binomial series gives,
\begin{align}
P_k(x)
&\propto
 \sum_{j=0}^{D-k}
\begin{pmatrix}
D-k\\ j
\end{pmatrix}
(-1)^{j} 
 I_{D,k+j}(x),
\end{align}
where
\begin{align}
I_{D,k}(x)
&\equiv
i^{D-1}
\int {d\xi\over 2\pi} 
{e^{i\xi^+(kx-1)}\over (\xi^+)^{D-1}}.
\end{align}
We then notice that for $kx>1$ the integral vanishes as 
one can close the integration contour in the 
upper complex plane without enclosing any singularity. 
For $1>kx$, on the other hand, the contour has to be closed in the lower complex 
plane and receives its contribution from  
the $D-1$-th order pole at $\xi=-i0$, i.e. 
\begin{align}
I_{D,k}(x)
&=
{  \theta(1-kx)
\over (D-2)!} 
\left(1- kx \right)^{D-2}.
\end{align}
Reorganizing the sum, and including a normalization 
constant, 
we then arrive at
\begin{align}
P_k(x)
&= 
{\cal N}
\sum_{j=k}^{{\rm min}\{D,[{1\over x}]\}}
\binom{D-k}{j-k}
(-1)^{j} 
\left(1-jx \right)^{D-2},
\end{align}
where $[x]$ is the largest integer $x_m$ with $x_m<x$, 
and the constraint reflects  
that $I_k(x)\propto\theta(1-kx)$.

{\it Normalization:---}To fix normalization, 
we notice that 
\begin{align}
\langle 1 \rangle
&=
\int_0^1 dx\,
 P_k(x)
 \nonumber\\
 &=
\frac{{\cal N}}{D-1} 
\sum_{j=k}^D
\binom{D-k}{j-k}
\frac{(-1)^j}{j} 
 \nonumber\\
 &=
\frac{(-1)^k(k-1)!(D-k)!}{D!(D-1)} 
{\cal N}.
\end{align}
That is, upon reorganizing factorials
\begin{align}
{\cal N}
 &=
(-1)^k
D(D-1)
\binom{D-1}{k-1}.
\end{align}

{\it First moment:---}Similarly, the average value
\begin{align}
\langle p_k \rangle
&=
\int_0^1 dx\,
x P_k(x)
 \nonumber\\
 &=
\frac{{\cal N}'}{D(D-1)}
\sum_{j=k}^D
\begin{pmatrix}
D-k\\ j-k
\end{pmatrix}
{(-1)^{j} \over j^2}
\nonumber\\
&=
{1\over D}
\left(
\psi(k) - \psi(D+1)
\right),
\end{align}
where $\psi(k)$ the digamma function.

\begin{table*}[t]
\centering
\begin{tabular}{c|ccccccccccccccccccc}
\hline
{\bf rank} & 1 & 2 & 3 & 4 & 5 & 6 & 7 & 8 & 9 & 10 & 11 & 12 & 13 & 14 & 15 & 16 & 17 & 18 & 19 \\
\hline
{\bf peak} & 0.492 & 0.471 & 0.470 & 0.466 & 0.470 & 0.472 & 0.468 & 0.463 & 0.464 & 0.466 & 0.468 
& 0.470 & 0.471 & 0.470 & 0.468 & 0.471 & 0.471 & 0.470 & 0.470 \\
\hline
\hline
{\bf circuit} & 1 & 2 & 3 & 4 & 5 & 6 & 7 & 8 & 9 & 10 & 11 & 12 & 13 & 14 & 15 & 16 & 17 & 18 & 19  \\
\hline
{\bf peak} & 0.459 & 0.439 & 0.493 & 0.525 & 0.443 & 0.493 & 0.497 & 0.471 & 0.465 & 0.483 & 0.453 
& 0.447 & 0.455 & 0.483 & 0.455 & 0.503 & 0.461 & 0.455 & 0.441  \\
\hline
\end{tabular}
\caption{
    Peak fidelity values obtained via maximum likelihood estimation under two averaging schemes: 
    (a) across the 20 circuits at fixed rank (top rows), 
    and (b) across the first 20 ranks within individual circuits (bottom rows), 
    with one scheme omitted per row for better readability. Results are based on top-$k$ 
    observed outcomes from the 12-qubit Sycamore dataset~\cite{App_Arute2019}.
}
\label{app:table}
\end{table*}

{\it Second moment:---}For the second 
moment
\begin{align}
\langle p^2_k \rangle
&=
\int_0^1 dx\,
x^2 P_k(x)
 \nonumber\\
 &=
\frac{2{\cal N}'}{(D+1)D(D-1)}  
\sum_{j=k}^D
\begin{pmatrix}
D-k\\ j-k
\end{pmatrix}
{(-1)^{j} \over j^3}
\nonumber\\
&=
\frac{1}{(D+1)D}  
\times
\nonumber\\
&\quad \times
\left(
[\psi(k)-\psi(D+1)]^2
+
\psi^{(1)}(k)-\psi^{(1)}(D+1)
\right),
\end{align}
or
\begin{align} 
\langle p^2_k\rangle
&=
{D\langle p_k\rangle^2 \over D+1}
+
{\psi^{(1)}(k)-\psi^{(1)}(D+1)
\over D(D+1)}.
\end{align}
Here $\psi(k)$ the digamma function and $\psi^{(1)}(k)$ its first derivative.

\section{Sycamore 12 analysis and scaling}

\begin{figure}[b!]
\centering
\includegraphics[width=8.0cm]{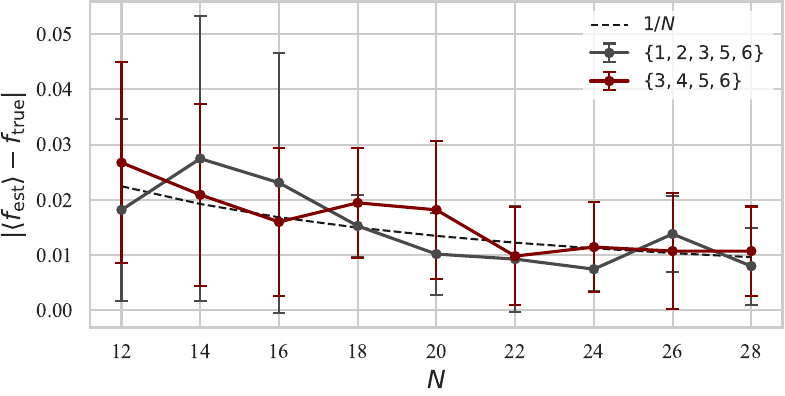}
\vspace{-.24cm}
\caption{
Scaling of fidelity estimation error with system size $N$ for $M=10$ Haar-random 
quantum states with true fidelity $f = 0.48$. Shown are two rank sets, 
$k = \{1,2,3,5,6 \}$ (grey) and $k = \{ 3,4,5,6 \}$ (red). Error bars indicate 
standard deviation across realizations.
}
\label{app:fig4}
\end{figure}

As discussed in the main text, we  consider two complementary averaging strategies 
for estimating circuit fidelity from the Sycamore 12 data: 
(a) combining likelihoods across circuits at a fixed rank, and (b) combining 
likelihoods across ranks within a single circuit. Table~\ref{app:table} summarizes 
the peak fidelity estimates from both averaging approaches
for the first 20 ranks (top rows) and first 20 circuits (bottom rows).
The fidelity estimates obtained by averaging across circuits at fixed rank lie in a 
narrow range, $f \approx 0.47-0.49$, indicating minimal rank dependence and 
supporting the assumption of a global, rank-independent noise model. In contrast, 
fidelity estimates from individual circuits, averaged over ranks, show broader 
variation, spanning a min-max range of $f \approx 0.44-0.52$, reflecting intrinsic 
differences in noise across circuit realizations. This spread is expected, 
as averaging over ranks retains circuit-specific noise characteristics and thus 
captures fidelity variations between circuits. Averaging across circuits 
at fixed rank, on the other hand, smooths out these realization-dependent fluctuations. 
To probe rank correlations, we repeated the analysis using sparse rank sampling and 
found that fidelity estimates remain practically unchanged compared to the 
full-rank case (deviations $< 0.01$).
Overall, the inferred fidelities cluster around $f \approx 0.45-0.50$, consistent 
with linear cross-entropy benchmarking (XEB) values reported in Ref.~\cite{App_Arute2019}. 

To further validate the scalability of our fidelity estimation approach, we simulate Haar-random 
quantum states for system sizes ranging from $N = 12$ to $N = 28$ qubits, each subject to homogeneous 
depolarizing noise of known strength. Fig.~\ref{app:fig4}  
shows the average estimation error $|\langle f_{\mathrm{est}} \rangle - f_{\mathrm{true}}|$ 
as a function of $N$ for $M=10$ realizations. The observed scaling is consistent 
with the expected $1/N$ behavior (dashed line), reflecting the statistical convergence 
of the likelihood estimator. 
As very low-rank statistics ($k = 1, 2$) exhibit larger fluctuations, we compare 
two rank sets, $k = \{1, 2, 3, 5, 6\}$ and $k = \{3, 4, 5, 6\}$, and find no significant 
differences in estimation accuracy. This supports the robustness of the method under practical 
rank selection constraints.

\section{Fidelity estimation under finite sampling}

The simulation validating the frequency-based fidelity estimator 
 consists of the following steps 
\begin{itemize}
    \item {\it Measurement Simulation:} For each rank $k$, 
    the count $n_k$ is sampled from a Poisson distribution with mean 
    $S p_k(f)$, where $p_k(f)$ reflects a mixture of structured (Haar) 
    and uniform noise.

    \item {\it Reranking of Counts:} The sampled counts are 
    sorted in descending order to generate an empirical ranking. 
    This step introduces realistic degeneracies and reshuffles 
    rank labels accordingly. Ties are resolved arbitrarily but 
    consistently.

    \item {\it Likelihood Maximization:} The fidelity parameter 
    $f$ is estimated by maximizing the log-likelihood over the 
    reranked counts.

    \item {\it Bisection for Minimal $S$:} To determine the 
    minimal number of shots required to reach a desired relative 
    error threshold in $f$, we perform a bisection search over $S$. 
    For each candidate value of $S$, we run multiple independent 
    realizations and analyze the resulting fidelity statistics. 
    Maximization is performed using bounded scalar optimization 
    over $f \in [0,1]$.
\end{itemize}

\end{appendix}


\end{document}